\RequirePackage{fix-cm}
\documentclass[twocolumn,epjc3]{svjour3}  
\smartqed  % flush right qed marks, e.g. at end of proof
\RequirePackage{graphicx}
\journalname{Eur. Phys. J. C}
\begin{document}

\title{A unique equation of state for the universe evolution from AdS$_5$ space-time}

\author{{M.M. Lapola}\thanksref{e1,addr1}
        \and
        {P.H.R.S. Moraes}\thanksref{e2,addr1,addr2}
        \and
        {W. de Paula}\thanksref{e3,addr1}
        \and
        {J. F. Jesus} \thanksref{e4,addr3}
        \and
        {R. Valentim} \thanksref{e5,addr4}
        \and
        {M. Malheiro} \thanksref{e6,addr1}}

\thankstext{e1}{e-mail: marcelo.lapola@gmail.com}
\thankstext{e2}{e-mail: moraes.phrs@gmail.com}
\thankstext{e3}{e-mail: wayne@ita.br}
\thankstext{e4}{e-mail: jfjesus@unesp.br}
\thankstext{e5}{e-mail: valentim.rodolfo@unifesp.br}
\thankstext{e6}{e-mail: malheiro@ita.br}

\institute{{ITA - Instituto Tecnol\'ogico de Aeron\'autica - Departamento de F\'isica,\\ 12228-900, S\~ao Jos\'e dos Campos, SP, Brazil} \label{addr1}
           \and
           {UNINA - Universit\`a degli Studi di Napoli Federico II - Dipartamento di Fisica\\\ Napoli I-80126, Italy}\label{addr2}
           \and
           {Universidade Estadual Paulista (Unesp), Campus Experimental de Itapeva - R. Geraldo Alckmin 519, 18409-010, Itapeva, SP, Brazil}\label{addr3}
           \and
           {Departamento de F\'{\i}sica, Instituto de Ci\^encias Ambientais, Qu\'{\i}micas e Farmac\^euticas (ICAQF), Universidade Federal de S\~ao Paulo - UNIFESP, Rua S\~ao Nicolau no.210, Centro, 09913-030, Diadema - SP, Brazil} \label{addr4}
}

\date{Received: date / Accepted: date}

\maketitle

\begin{abstract}
{We apply the Induced Matter Model to a five-dimensional metric. For the case with null cosmological constant, we obtain a solution able to describe the radiation-dominated era of the universe. The positive $\Lambda$ case yields a bounce cosmological model. In the negative five-dimensional cosmological constant case, the scale factor is obtained as $a(t)\sim\sqrt[]{\sinh t}$, which is able to describe not only the late-time cosmic acceleration but also the non-accelerated stages of the cosmic expansion in a continuous form.  This solution together with the extra-dimensional scale factor solution yields the material content of the model to be remarkably related through an equation of state analogous to the renowned MIT bag model equation of state for quark matter $p=(\rho-4B)/3$. In our case, $\rho=\rho_m+B$, with $\rho_m$ being the energy density of relativistic and non-relativistic matter and $B=\vert\Lambda\vert/16\pi$ represents the bag energy constant, which plays the role of the dark energy in the four-dimensional universe, with $\Lambda$ being the cosmological constant of the AdS$_5$ space-time. Our model satisfactorily fits the observational data for the low redshift sample of the experimental measurement of the Hubble parameter, which resulted in $H_0=72.2^{+5.3}_{-5.5}$km s$^{-1}$ Mpc$^{-1}$.}

\end{abstract}

\section{Introduction}
\label{intro}

Since the late 90's of the last century, a lot of efforts have been made to describe the observable universe as a brane embedded in a higher dimensional space \cite{sundrum/1999}-\cite{flanagan/2000}. Some results obtained from such a set up for the universe are remarkable. Braneworld models of dark energy were recently presented in References \cite{jawad/2015}-\cite{rani/2016}. In Reference \cite{sahni/2005}, the possibility of the $\Lambda$CDM cosmological model be a braneworld model in disguise was investigated.  In the astrophysics of compact objects context, braneworld models are able to predict some deviations from standard General Relativity (GR) outcomes and get in touch with some peculiar observations \cite{germani/2001}-\cite{lugones/2017}. To be in touch with recent literature on braneworld models applications, we suggest References \cite{prasetyo/2018}-\cite{barbosa-cendejas/2014}.

The braneworld scenario was originally proposed as an alternative to the hierarchy problem, as it can be checked, for instance, in References \cite{yang/2012}-\cite{das/2008}. The concept of extra dimensions has also been used in attempts to unify the four fundamental forces of nature \cite{hall/2002}-\cite{appelquist/1984}.

Not only on braneworld scenarios lie the extra dimensional universe configurations. There are also the renowned Kaluza-Klein (KK) models \cite{visser/1985}-\cite{hohm/2013}. The relic density of KK dark matter in universal extra dimensions was calculated \cite{kong/2006}. The virtual effects of KK states on Higgs physics in universal extra dimensional models were examined \cite{petriello/2002}. F. Darabi and P.S. Wesson introduced a generalized gravitational conformal invariance in the context of non-compactified five-dimensional (5D) KK theory \cite{darabi/2002}. In astrophysics, stability of strange stars in extra dimensions has been investigated  recently \cite{malheiro/2019}.

P.S. Wesson, has contributed very significantly to KK cosmology as well as for the interpretation of the extra dimension \cite{wesson/1992}-\cite{wesson/1992b}. Together with collaborators, Wesson has also investigated the effective properties of matter in KK theory \cite{liu/1994}, outlined a Machian interpretation of KK gravity \cite{mashhoon/1994}, applied some classical tests to the theory \cite{kalligas/1995} and derived the referred equation of motion \cite{wesson/1995}.

The outcomes of some more recent articles by Wesson et al. on five-dimensional (5D) universe can be appreciated in the following. J.M. Overduin et al. have used measurements of geodesic precession from Gravity Probe B experiment and constrained possible departures from Einstein's GR for a spinning test body in KK theory \cite{overduin/2013}. C. Zhang and collaborators have used Wetterich's parameterization equation of state (EoS) to obtain cosmological solutions in a 5D Ricci-flat Universe \cite{zhang/2006}. In \cite{seahra/2002}, some relations for the embedding of spatially flat Friedmann-Lema\^itre-Robertson-Walker (FLRW) cosmological models in flat KK manifolds were presented.

The cosmological constant problem, namely, the huge discrepancy between theoretical and observed values of the cosmological constant in standard $\Lambda$CDM cosmology, was investigated in KK gravity by P.S. Wesson and H. Liu \cite{wesson/2001}. F. Darabi et al. have derived a quantum cosmology from KK theory with non-compactified extra dimension \cite{darabi/2000}. In \cite{wesson/2000}, Wesson et al. have obtained an exact solution of the 5D field equations that describes a shock wave moving in time and extra KK coordinate. Such a solution suggested that the four-dimensional (4D) big bang was a 5D shock wave.

Particularly regarding the interpretation of the extra dimension in the 4D observable universe, Wesson has proposed the so-called Induced Matter Model (IMM), which can be appreciated in \cite{liu/1994},\cite{moraes/2015}-\cite{fukui/2001}. It consists of the following concept. The KK field equations read 

\begin{equation}\label{i1}
G_{AB}=0,
\end{equation}
with $G_{AB}$ being the Einstein tensor and the indices $A,B$ run from $0$ to $4$. From Eq.(1), it can be seen that the KK field equations depend only on the 5D metric $g_{AB}$. The Wesson's idea consists of collecting in Eq.(1) the terms that depend on the extra coordinate and make them play the role of an induced energy-momentum tensor in 4D. Further applications of the IMM can be appreciated in Refs.\cite{ponce_de_leon/2010}-\cite{halpern/2000}.

In the present article we intend to develop - by meaning of Wesson's model - and investigate the Friedmann-like equations derived from a 5D metric. The field equations will be taken as Eq.(1) in the presence of a 5D cosmological constant $\Lambda$, that is \cite{sahni/2003}

\begin{equation}\label{ie}
G_{AB}+\Lambda g_{AB}=0.
\end{equation}

We will be particularly concerned with the role of the extra-dimension scale factor in the metric \cite{mm/2012}-\cite{la_camera/2010}

\begin{equation}\label{i3}
ds^{2}=dt^{2}-a(t)^{2}[dr^{2}+r^{2}(d\theta^{2}+\sin^{2}\theta d\phi^{2})]-\xi(t)^{2}dl^{2}.
\end{equation}
In Eq.(\ref{i3}), $a(t)$ is the scale factor of the observable universe and $\xi(t)$ is the extra-dimension scale factor. Moreover, we are assuming the spatial curvature of the universe to be null, in accordance with recent observational data on the fluctuations of temperature of the cosmic microwave background radiation \cite{hinshaw/2013}. Still in (\ref{i3}), $t$ is the time coordinate, $r,\theta$ and $\phi$ are the polar spherical coordinates and $l$ is the extra spatial coordinate. Throughout the article, natural units will be assumed, unless otherwise advised.

In the present article, we shall investigate the cosmological solutions obtained from the substitution of (\ref{i3}) in (\ref{i1}) and in (\ref{ie}).  We will search for being in touch with recent cosmological observational data, which shall naturally constrain the extra dimensional features of the model. 

\section{4D dynamics from 5D empty space}

In the present section we will substitute Eq.(\ref{i3}) in Eqs.(\ref{i1}) and (\ref{ie}). For all cases we will consider that matter in the 4D observable universe is a manifestation of a 5D universe devoid of matter, through the application of the IMM. That is to say that the terms on the 5D Einstein tensor for (\ref{i3}) which somehow depend on the extra coordinate will ``be moved'' to the {\it rhs} of Eqs.(\ref{i1}) and (\ref{ie}) to play the role of an induced energy-momentum tensor.

Throughout the whole article, the energy-momentum tensor of a perfect fluid will be assumed, that is, $T_{A}^{B}=\mbox{diag}(\rho,-p,-p,-p,0)$, with $\rho$ being the matter-energy density and $p$ the pressure of the universe. Note that $T_4^4=0$ since we will consider, such as in braneworld models, that matter is restricted to the 4D universe.

\subsection{Field equations without cosmological constant}\label{ss:fewcc}

The non-null components of the Einstein tensor obtained when substituting metric (3) in field equations (1) read:
\begin{equation}
G_{0}^{0}=3\left[\left(\frac{\dot{a}}{a}\right)^{2}+\left(\frac{\dot{a}}{a}\right)\left(\frac{\dot{\xi}}{\xi}\right)\right],
\label{t4}
\end{equation}
\begin{equation}
G_{1}^{1}=G_{2}^{2}=G_{3}^{3}=\left(\frac{\dot{a}}{a}\right)^{2}+2\left(\frac{\dot{a}}{a}\right)+2\left(\frac{\ddot{a}}{a}\right)+\frac{\ddot{\xi}}{\xi},
\end{equation}
\begin{equation}
G_{4}^{4}=3\left[\left(\frac{\dot{a}}{a}\right)^{2}+\left(\frac{\ddot{a}}{a}\right)\right],
\label{t6}
\end{equation}
where dots indicate time derivatives.

By applying the IMM in such components, one has

\begin{eqnarray}
\rho= - \frac{3}{8\pi}\frac{\dot{a}}{a}\frac{\dot{\xi}}{\xi}, \label{fewcc1}\\
p= \frac{1}{4\pi}\left(\frac{\dot{a}}{a}\frac{\dot{\xi}}{\xi}+\frac{1}{2}\frac{\ddot{\xi}}{\xi}\right)\label{fewcc2},
\end{eqnarray}
and from $G_{44}=0$ we also obtain the constrain equation

\begin{equation}
\left(\frac{\dot{a}}{a}\right)^{2}+\frac{\ddot{a}}{a}=0. \label{fewcc3}
\end{equation}

By solving Eq.(\ref{fewcc3}), we have

\begin{equation}\label{scalefactor}
a(t) = c_{1}\sqrt{t}.
\end{equation}
Throughout the article, $c_i$, with $i=1,2,3,...$, are constants.

It is worth to remark that Eq.(10) is in agreement with a radiation-dominated universe, since in standard cosmology, $a\sim t^{1/2}$ occurs exactly for such a stage of the universe evolution \cite{ryden/2003}. Remarkably, when Kaluza developed his extradimensional theory of gravity, today called KK gravity, his intent was to describe from $G_{AB}=0$ uniquely, both 4D Einstein's field equations with matter and Maxwell's equations for electromagnetism, as it can be checked in \cite{overduin/1997}.

We can substitute (10) in the $G_{00}$ component of Eq.(1) for metric (3) and derive the solution for $\xi(t)$ in

\begin{equation}\label{00}
\frac{\dot{a}}{a}+\frac{\dot{\xi}}{\xi}=0.
\end{equation}
The result is
\begin{equation}\label{ecscalefactor}
\xi(t) = \frac{c_2}{\sqrt{t}}.
\end{equation}
It can be verified that Eqs. (10) and (12) are solutions of Eq.(1).

It is interesting to remark that the solution obtained for $\xi(t)$ may indicate a compactification of the extra coordinate as time passes by. This can be clearly verified by deriving the referred Hubble parameter $H_l=\dot{\xi}/\xi$, which reads 
\begin{equation}\label{echubble}
H_l(t) = -\frac{1}{2\, t},
\end{equation}
and a negative Hubble parameter would indicate compactification rather than expansion of the referred space.

Solutions (10) and (12) when substituted in (7) and (8) yield, respectively,
\begin{equation}\label{rho1}
\rho(t) = \frac{3}{32\pi \, t^2},
\end{equation}
\begin{equation}\label{p1}
p(t) = \frac{1}{32\pi \, t^2}.
\end{equation}
We can see from Eqs.(14) and (15) that $\rho$ and $p$, in this model, have a quadratic term on $t$. Such a behaviour can also be seen in braneworld models \cite{sahni/2003,keresztes/2007}.

We can also note that, remarkably, by dividing (15) by (14), one has $\omega=p/\rho=1/3$, which is the EoS parameter of a radiation-dominated universe \cite{ryden/2003}. This result can also be verified in the Friedmann-like equations (7)-(8).

\subsection{Field equations with cosmological constant}\label{ss:fewcc2}

\subsubsection{Case I: $\Lambda>0$}\label{sss:lg0}

Let us now work with Eq.(2). By substituting metric (3) in (2), we can write, through the IMM application, the following Friedmann-like equations:

\begin{eqnarray}
\rho = - \frac{3}{8\pi}\left(\frac{\dot{a}}{a}\frac{\dot{\xi}}{\xi}+\frac{\Lambda}{3}\right), \label{t21}\\
p = \frac{1}{8\pi}\left(2\frac{\dot{a}}{a}\frac{\dot{\xi}}{\xi}+\frac{\ddot{\xi}}{\xi}+\Lambda\right), \label{t22}\\
\left(\frac{\dot{a}}{a}\right)^{2}+\frac{\ddot{a}}{a}=-\frac{\Lambda}{3}. \label{t23}
\end{eqnarray}

Eq.(18) can be solved for the scale factor, yielding
\begin{equation}\label{t24}
a(t) = c_3\sqrt{\Big\vert\sin\left(\sqrt{\frac{2}{3}\Lambda} \, t\right)\Big\vert}.
\end{equation}

\begin{figure*}[]
\centering
\includegraphics[width=105mm]{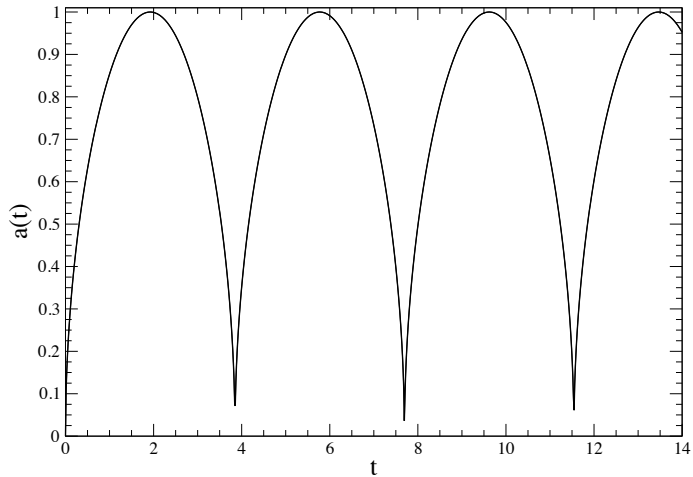}
\caption{Evolution of the scale factor as a function of time in natural units, for $c_3=\Lambda=1$.}
\label{fig1}
\end{figure*}

The evolution of the scale factor (19) in time can be appreciated in Fig.\ref{fig1}.

By analysing Fig.\ref{fig1} we are led to conclude that a positive 5D cosmological constant yields a cyclic or bouncing universe \cite{steinhardt/2002}-\cite{battefeld/2015}. 

In possession of Eq.(19), we can use the non-null components of Eq.(2) to write

\begin{equation}\label{lp1}
\xi(t) = c_4\frac{\big\vert\cos\left(\sqrt{\frac{2}{3}\Lambda} \, t\right)\big\vert}{\sqrt{\big\vert \sin \left(\sqrt{\frac{2}{3}\Lambda} \, t\right) \big\vert}}.
\end{equation}

\begin{figure*}[h!]
\centering
\includegraphics[width=105mm]{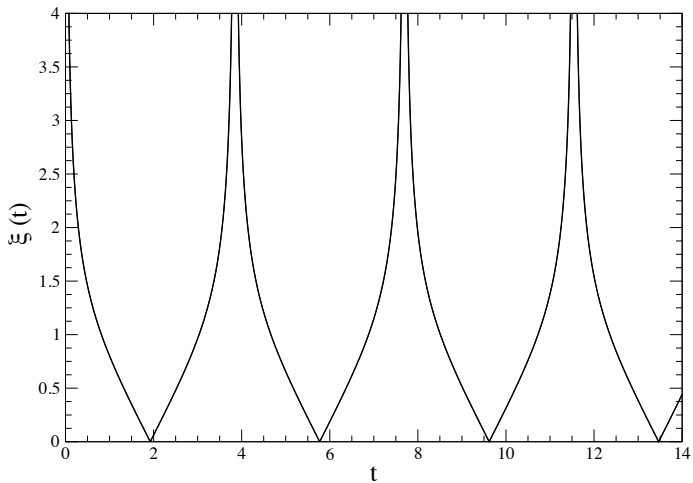}
\caption{Evolution of the extra-dimension scale factor as a function of time in natural units, for $c_4=\Lambda=1$.}
\label{fig2}
\end{figure*}

From Fig.\ref{fig2}, we can see that $\xi$ completes each cycle in the same time scale as $a$ does. We can also see that, by keeping in mind that $a=1$ at present, the length scale of the extra dimension is minimum today, which could justify the absence of evidences of extra dimensions in the Large Hadron Collider \cite{chatrchyan/2012}-\cite{datta/2013}.

From (19) and (20), we can write the explicit solutions for $\rho(t)$ and $p(t)$ as
\begin{eqnarray}
\rho(t) = \frac{\Lambda}{16 \pi } \cot^2\left(\sqrt{\frac{2}{3}\Lambda} \, t\right), \\
p(t) = \frac{\Lambda}{48 \pi }\left[\cot^2\left(\sqrt{\frac{2}{3}}\Lambda \, t \right)+4\right].
\end{eqnarray}
\label{sec:1}

Although bouncing models have their importance specially because they evade the Big-Bang singularity, we should discard the present model due to the impossibility of predicting the late-time accelerated expansion regime of the universe \cite{riess/1998,perlmutter/1999} from Eq.(19).

\subsubsection{Case II: $\Lambda<0$}\label{sss:ll0}

Following the same approach of the previous section now for $\Lambda<0$, we obtain the scale factors as

\begin{eqnarray}\label{t25}
a(t) &=& c_5\sqrt{\sinh\left(\sqrt{\frac{2}{3}\vert\Lambda\vert} \, t \right)} \, , \\
\xi(t) &=& c_6 \frac{\cosh \left(\sqrt{\frac{2}{3} \vert\Lambda\vert} \, t \right)}{\sqrt{\sinh \left(\sqrt{\frac{2}{3} \vert\Lambda\vert }  \, t \right)}}. \label{5DscaleNegConstant}
\end{eqnarray}

The evolution of those scale factors can be appreciated in Figures \ref{fig3}-\ref{fig4} below.

\begin{figure*}[h!]
\centering
\includegraphics[width=105mm]{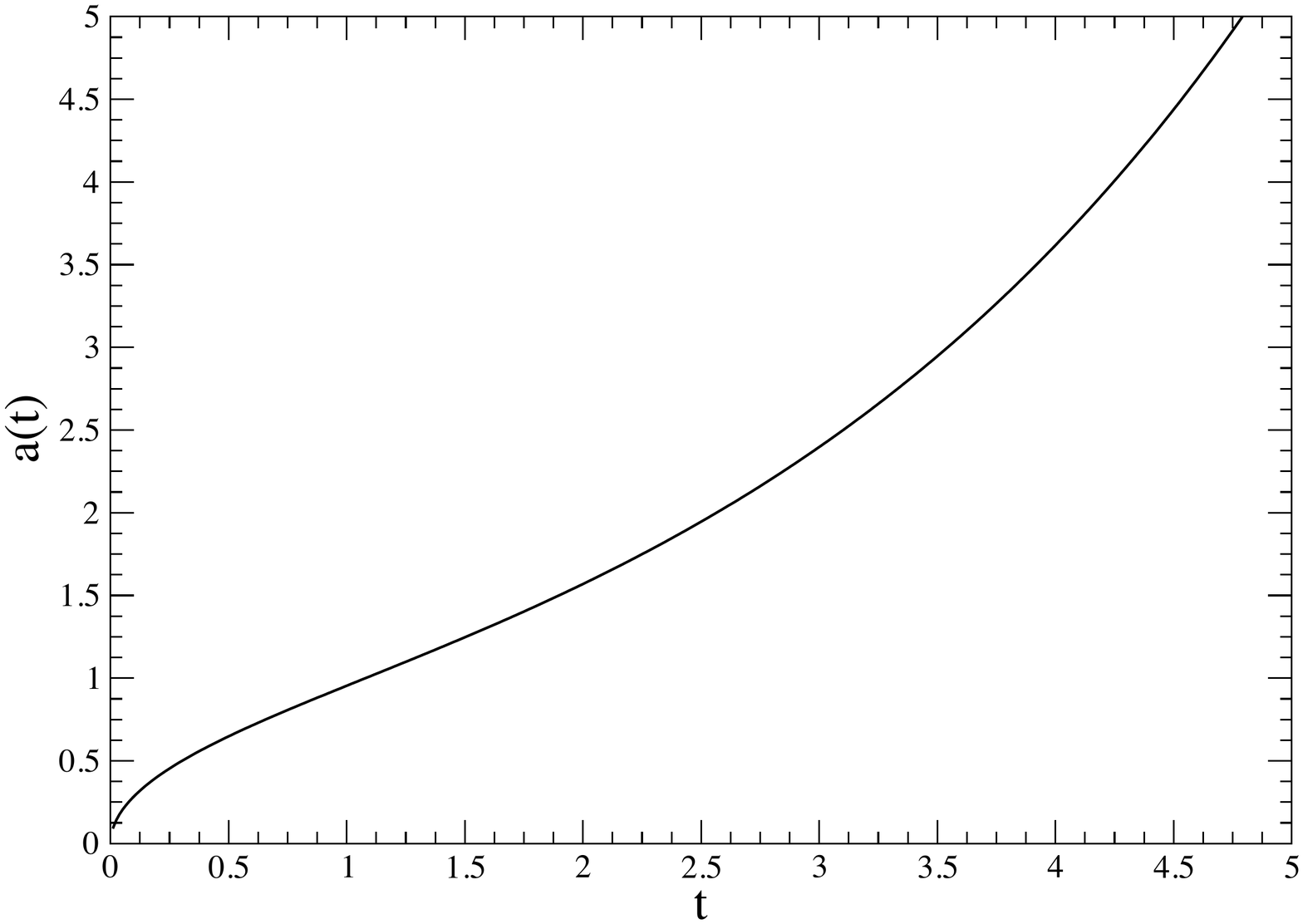}
\caption{Evolution of the scale factor as a function of time in natural units, for $c_5=1$ and $\Lambda=-1$.}
\label{fig3}
\end{figure*}

\begin{figure*}[h!]
\centering
\includegraphics[width=105mm]{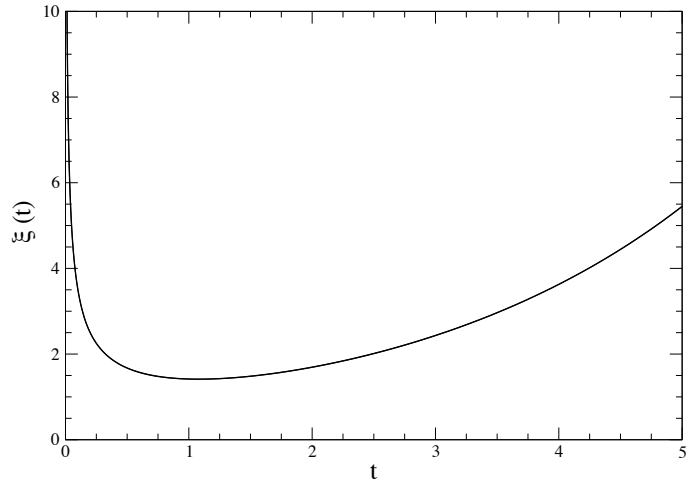}
\caption{Evolution of the extra-dimension scale factor as a function of time in natural units, for $c_6=1$ and $\Lambda=-1$.}
\label{fig4}
\end{figure*}

We can see from Fig.\ref{fig3} that $a(t)$ assumes an exponential behaviour as time grows, which may be an indication of the recent cosmic acceleration \cite{riess/1998,perlmutter/1999}. This will be clarified in Fig.\ref{fig5} below.

From Fig.\ref{fig4}, we can see that the extra dimension apparently is large for the primordial stages of the universe. Then, it naturally suffers a process of compactification, assuming its minimum value for $t\sim1$. Then, it maximizes its length scale once again.

It is possible to derive a relation between the scale factors $a(t)$ and $\xi(t)$. Starting from (4) and (6) for $\Lambda<0$ we obtain the system of equations
\begin{eqnarray}
\frac{\dot{a}}{a}\frac{\dot{\xi}}{\xi}+\left(\frac{\dot{a}}{a}\right)^{2} &=&   \frac{\Lambda}{3}
\label{t29} \\
\left(\frac{\ddot{a}}{\dot{a}}\right)+\left(\frac{\dot{a}}{a}\right)^{2} &=&  \frac{\Lambda}{3}
\label{t30}
\end{eqnarray}
Subtracting (27) from (28), leads to

\begin{equation}
\frac{\dot{\xi}}{\xi}=\frac{\ddot{a}}{\dot{a}} \, .
\end{equation}
Thus, we obtain a relation between the extra-dimension scale factor $\xi$ and the time derivative of $a$ as

\begin{equation}
\xi=K\dot{a},
\end{equation}
 with constant $K$. Therefore,

\begin{equation}
\frac{\xi}{a}=KH=\frac{c_6}{c_5}\sqrt{\frac{6}{\vert\Lambda\vert}}H,
\label{t39}
\end{equation}
where $H=H(t)=\frac{\dot{a}}{a}$ is the Hubble parameter.

The solutions for the induced matter content read

\begin{eqnarray}
\rho(t) &=& \frac{\vert\Lambda\vert}{16 \pi}  \coth ^2\left(\sqrt{\frac{2}{3}\vert\Lambda\vert} \, t \right) \, ,\label{rholl0} \\
p(t) &=& \frac{\vert\Lambda\vert} {48 \pi } \left[\coth ^2\left(\sqrt{\frac{2}{3}\vert\Lambda\vert}  \, t \right)-4\right] \, ,\label{pll0}
\end{eqnarray}
and the induced density can be rewritten as

\begin{equation}
\rho=\frac{\vert\Lambda\vert}{16\pi}\left(\frac{c_5}{c_6}\frac{ \, \xi}{ \, a}\right)^{2}
\end{equation}
or

\begin{equation}
\rho=\frac{3H^{2}}{8\pi},
\end{equation}
where

\begin{equation}
%\begin{cases}
%H(t)=\sqrt{\frac{\Lambda}{6}} \, \cot\left(\sqrt{\frac{2}{3}\Lambda} \, t \right), & \Lambda>0\\
H(t)=\sqrt{\frac{\left|\Lambda\right|}{6}}\, \coth\left(\sqrt{\frac{2}{3}\left|\Lambda\right|} \, t\right).
%\end{cases}
\end{equation}

The Hubble parameter has its evolution in time shown in Figure \ref{fig5}.

\begin{figure*}[h!]
\centering
\includegraphics[width=105mm]{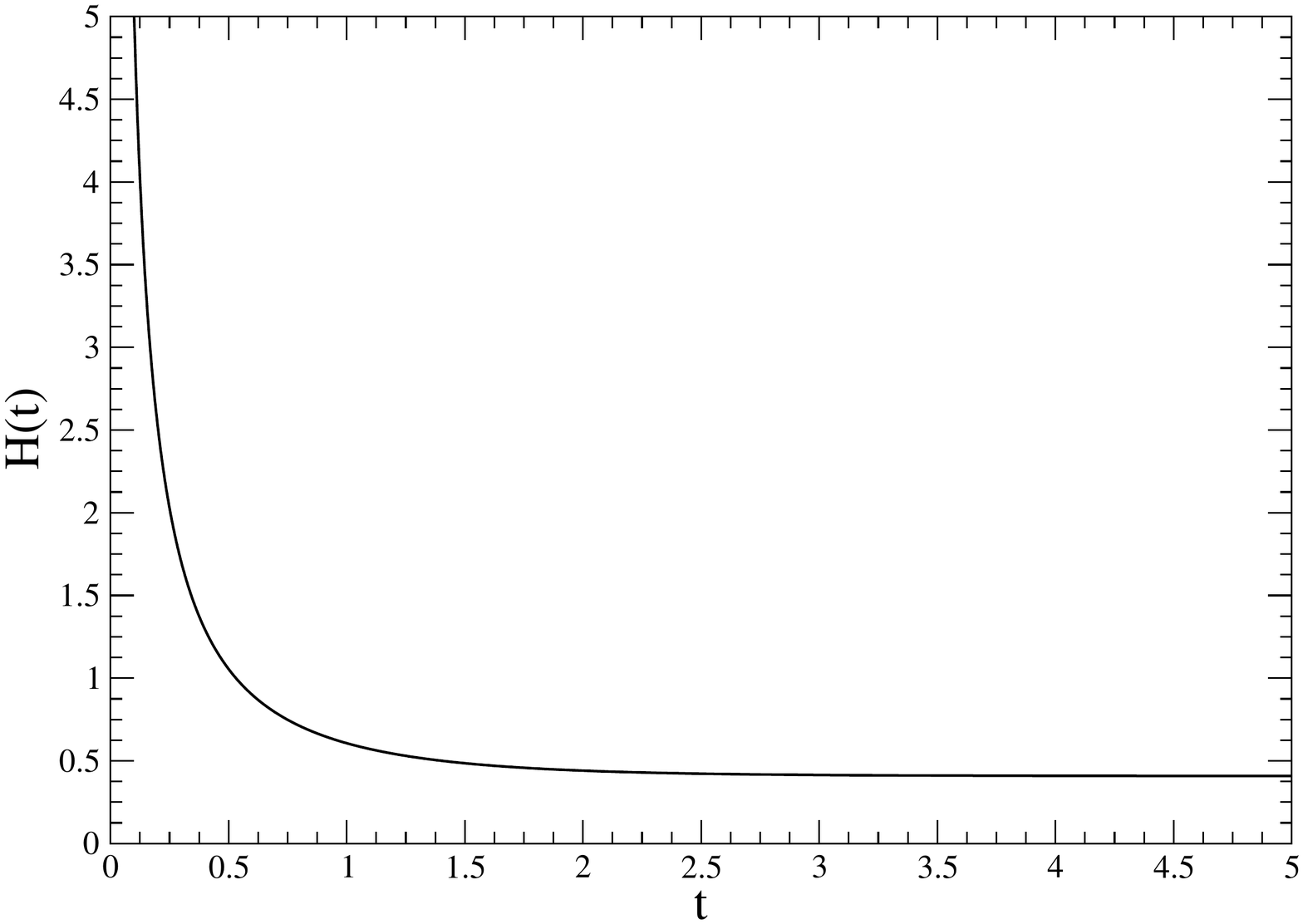}
\caption{Evolution of the Hubble parameter as a function of time in natural units, for $\Lambda=-1$.}
\label{fig5}
\end{figure*}

We see from Fig.\ref{fig5} that the predicted Hubble parameter starts evolving as $\sim1/t$, which is, indeed, expected from standard model predictions \cite{ryden/2003}. After a period of time, $H(t)\sim\ constant$. It is known that an exponential scale factor describes the cosmic acceleration. From the definition of the Hubble parameter, an exponential scale factor yields a constant Hubble parameter. In this way, the constant behaviour that $H(t)$ assumes for high  values of time is an indication of the recent cosmic acceleration.

\section{Bag model-like equation of state for the universe evolution and the deceleration parameter}\label{sec: eos}

In this section we will deeper investigate the solutions obtained in the previous section for the negative 5D cosmological constant case. We will show that the induced matter-energy density and pressure can be related through a bag model-like EoS. We will also derive, from the scale factor solution, the deceleration parameter of the model.

\subsection{Unified equation of state for the universe evolution}

Considering the cases in which $\Lambda\neq0$, the expressions for the density and pressure can be remarkably written in a unified form as

\begin{equation}
p=\frac{\left(\rho\pm4B\right)}{3}, 
\label{t36}
\end{equation}
where the constant $B=\frac{|\Lambda|}{16\pi}$  and the positive sign stands for $\Lambda>0$ while the negative sign for $\Lambda<0$. Eq.(\ref{t36}) remarkably resembles the MIT bag model equation of state (check, for instance \cite{maieron/2004,nicotra/2006,dkmkmr/2019}), for which $B$ is the so-called bag constant. Naturally we are not claiming that the universe is made of quarks confined inside a bag, but that the EoS for the universe has the same mathematical form. The bag constant here is in fact the bulk energy density necessary to create the vacuum in the flat 5D space, and in this sense, it plays the same role of the bag constant in the MIT model, that is, the energy density necessary to create a bag in the QCD vacuum. As we will see, for the universe evolution given by this EoS, the constant $B$ will be identified with the dark energy in the 4D universe.

For the case of cosmological interest, namely the case $\Lambda<0$, we can separate the density in matter-radiation and dark energy components, so that

\begin{eqnarray}
\label{t541}
\rho &=& \rho_{m}+\rho_{\Lambda} \, ,  \\
\rho_{m} &=& \frac{|\Lambda|}{16\pi} {cosech}^{2}\left(\sqrt{\frac{2}{3}|\Lambda|}t\right) \, , \\
\rho_{\Lambda} &=& \frac{|\Lambda|}{16\pi} \, .
\label{t54}
\end{eqnarray}
We can see from (35) and (38)  that the bag energy constant $B$ in this bag model-like unified EoS for the universe plays the role of dark energy in the 4D universe.

One can write the EoS parameter as
%\begin{equation}
%p=\frac{\left(\rho-4B\right)}{3} \, ,
%\label{t41}
%\end{equation}
%which can be rewritten as
%\begin{equation}
%p=\frac{\rho-4(\rho-\rho_{m})}{3}=-\rho+\frac{4}{3}\rho_{m} \, .
%\end{equation}
%Dividing both sides of this equation by $\rho$ we obtain 
\begin{equation}
\omega=-1+\frac{4}{3}{sech}^{2}\left(\sqrt{\frac{2}{3}|\Lambda|}t\right)
\end{equation}

whose evolution in time can be appreciated in Fig.6.

\begin{figure*}[h!]
\centering
\includegraphics[width=105mm]{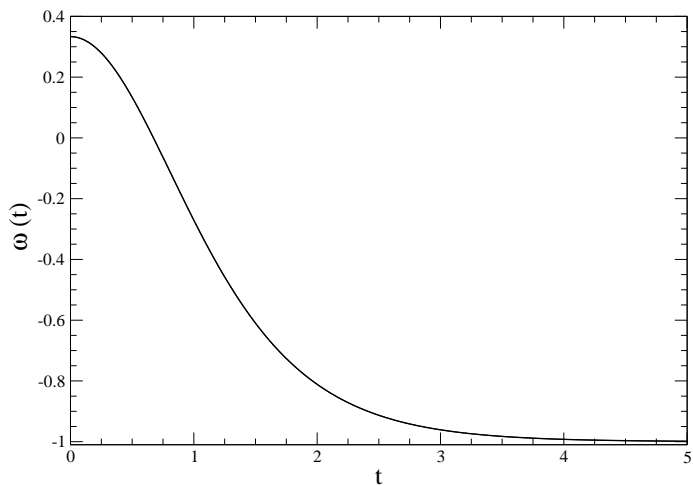}
\caption{Evolution of the equation of state parameter as a function of time in natural units, for $\Lambda=-1$.}
\label{fig6}
\end{figure*}

The model prediction for the evolution of the EoS parameter in time, according to Fig.\ref{fig6}, is remarkable. One can note that for small values of time,  $\omega\sim1/3$. According to standard model (as it was mentioned above), the primordial value of $\omega$ is indeed $1/3$, as the primordial universe dynamics is dominated by radiation, such that $p=\rho/3$ \cite{ryden/2003}. As the universe expands and cools down, it allows pressureless matter to be formed. This stage represents the matter-dominated stage of the universe, for which $\omega\sim0$, that is also depicted in Fig.\ref{fig6}. Last, but definitely not least, Fig.\ref{fig6} indicates that for high values of time, $\omega\sim-1$. According to recent observations on the cosmic microwave background radiation temperature fluctuations, $\omega=-1.073^{+0.090}_{-0.089}$ \cite{hinshaw/2013}. This negative pressure fluid is the responsible for the cosmic acceleration in standard model. Therefore, our present approach has also revealed a dominant negative pressure fluid for high values of time, but remarkably, it has also predicted other stages of the universe evolution, named radiation and matter-dominated eras, in a continuous and analytical form.

It is important to show that the expression (36) for the density satisfies the continuity equation in 4D. Starting from the continuity equation
\begin{equation}
\dot{\rho}+3\frac{\dot{a}}{a}(\rho+p)=0, \,
\end{equation}
substituting (35) and integrating on both sides leads to
\begin{equation}
\rho(t) = \frac{c_{7}}{a(t)^{4}} + B.
\end{equation}
Comparing the last expression with Eqs. (36) and (38) and using the expression (23) for $a(t)$, we obtain $c_{7}={c_{5}}^{4} \frac{|\Lambda|}{16\pi}$, that proves that the continuity equation in 4D is satisfied in our model.

\subsection{The deceleration parameter}

The deceleration parameter is defined as 
\begin{equation}
q(t) =-\frac{\ddot{a} \, a}{\dot{a}^{2}},%=-\frac{\dot{\xi}(t)}{\xi(t)}\frac{a(t)}{\dot{a}(t)}=-\frac{H_{l}(t)}{H(t)} \, ,\label{deceleration}
\end{equation}
so that $q>0$ indicates a decelerated expansion and $q<0$ indicates an accelerated expansion.

In the present model, it can be show that

\begin{equation}
q=\frac{\dot{\xi}}{\xi}\frac{a}{\dot{a}}=-\frac{H_{l}}{H}.   
\end{equation}
Therefore, remarkably the deceleration factor in our model is the negative ratio between the Hubble parameter of the extra-dimension scale factor and the Hubble parameter in 4D.

Explicitly, the deceleration parameter for $\Lambda<0$ reads

\begin{equation}
%q(t) &=& 1+2 \tan^2\left(\sqrt{\frac{2}{3}\Lambda} \, t \right) \, , \\
q(t) = 1-2 \tanh^2\left(\sqrt{\frac{2}{3}|\Lambda|}\, t \right). \, 
\label{t48}
\end{equation}

%Now, let us relate the deceleration factor with $\rho_{m}$ and $\rho_{\Lambda}$ for the case of cosmological interest $(\Lambda<0)$.

%From \eqref{(31)} and \eqref{(38)}, we can rewrite the density as

%\begin{equation}
%\rho(t)=\frac{\vert\Lambda\vert}{16\pi}\left(\frac{c_5}{c_6}\frac{\xi}{a}\right)^{2}\,, 
%\end{equation}
%so that
%\begin{equation}
%\rho(t)=\frac{3}{8\pi}H^{2}.
%\end{equation}

\section{Cosmological parameters in terms of redshift and the observational analysis}
\label{Hubble}

With the purpose of confronting our solutions with observational data, we will study the behavior of the Hubble parameter and other cosmological parameters in terms of the redshift rather than of time. We will concentrate our attention in the case $\Lambda<0$.

Taking into account the scale factor obtained in (23), the redshift can be written as

\begin{equation}\label{redshiftLambdaPositive}
z(t) = -1 + \frac{1}{c_5\sqrt{\sinh\left(\sqrt{\frac{2}{3}\vert\Lambda\vert} \, t \right)}} .
\end{equation}

The Hubble parameter is then expressed in terms of redshift as follows

\begin{equation}
H(z)= \sqrt{\frac{\vert\Lambda\vert}{6}}\coth \left[arcsinh\left[\frac{1}{c_5(1+z)}\right]^{2}\right\} \, .
\end{equation}

The above equation gives a relation between the Hubble constant, the cosmological constant and the integration constant $c_5$ as
\begin{equation}
H_0=\sqrt{\frac{\vert\Lambda\vert}{6}}\coth\left[arcsinh\left(\frac{1}{c_5}\right)^{2}\right] \, .
\label{H0c5L}
\end{equation}
%\begin{equation}
%\frac{|\Lambda|}{6 H_0^2} = %\left(\frac{1}{\coth \left( %\arcsinh \left(c_5^{-2}\right) %\right)}\right)^2
%\end{equation}

\subsection{Observational constraints}

Hubble parameter data as a function of redshift yields one of the most straightforward cosmological tests today. It consists on constraining the cosmological models with values of the expansion rate as a function of redshift. It is even more interesting when the Hubble parameter data come from estimates of differential ages of objects at high redshifts, because it is inferred from astrophysical observations alone, not depending on any background cosmological models (check References \cite{SternEtAl10,zt}).

The data we use here comes from the 51 $H(z)$ data compilation from Maga\~na {\it et al.} \cite{Magana2018}. This compilation consists of 20 clustering (from Baryon Acoustic Oscillations and Luminous Red Galaxies) and 31 differential age $H(z)$ data. 

We choose to work here only with the 31 differential age $H(z)$ data\footnote{Marked as ``DA'' in Table 1 of Ref. \cite{Magana2018}.}, because it does not depend on any background cosmological model. The age estimates depend only on models of chemical evolution of objects at high redshifts. $H(z)$ estimates from clustering like Baryon Acoustic Oscillations usually assume a standard cosmological model in order to obtain the data from surveys.

In all analyses here, we have written a $\chi^2$ function for parameters, with the likelihood given by ${\mathcal L}\propto e^{-\chi^2/2}$. The $\chi^2$ function for $H(z)$ data is given by the following:
\begin{equation}
\chi^2_H = \sum_{i = 1}^{31}\frac{{\left[ H_{obs,i} - H(z_i,\mathbf{s})\right] }^{2}}{\sigma^{2}_{H_i,obs}} ,
\label{chi2H}
\end{equation}
where $\mathbf{s}$ is the parameter vector, which we choose to be $\mathbf{s}=(c_5,H_0)$. $\Lambda$ can be related to these parameters through Eq.(\ref{H0c5L}).

In Figure \ref{Hz31} below, we can see the 31 $H(z)$ data used here and the best fit $H(z)$ we have found by minimizing $\chi^2_H$.

\begin{figure*}[h!]
\centering
\includegraphics[width=105mm]{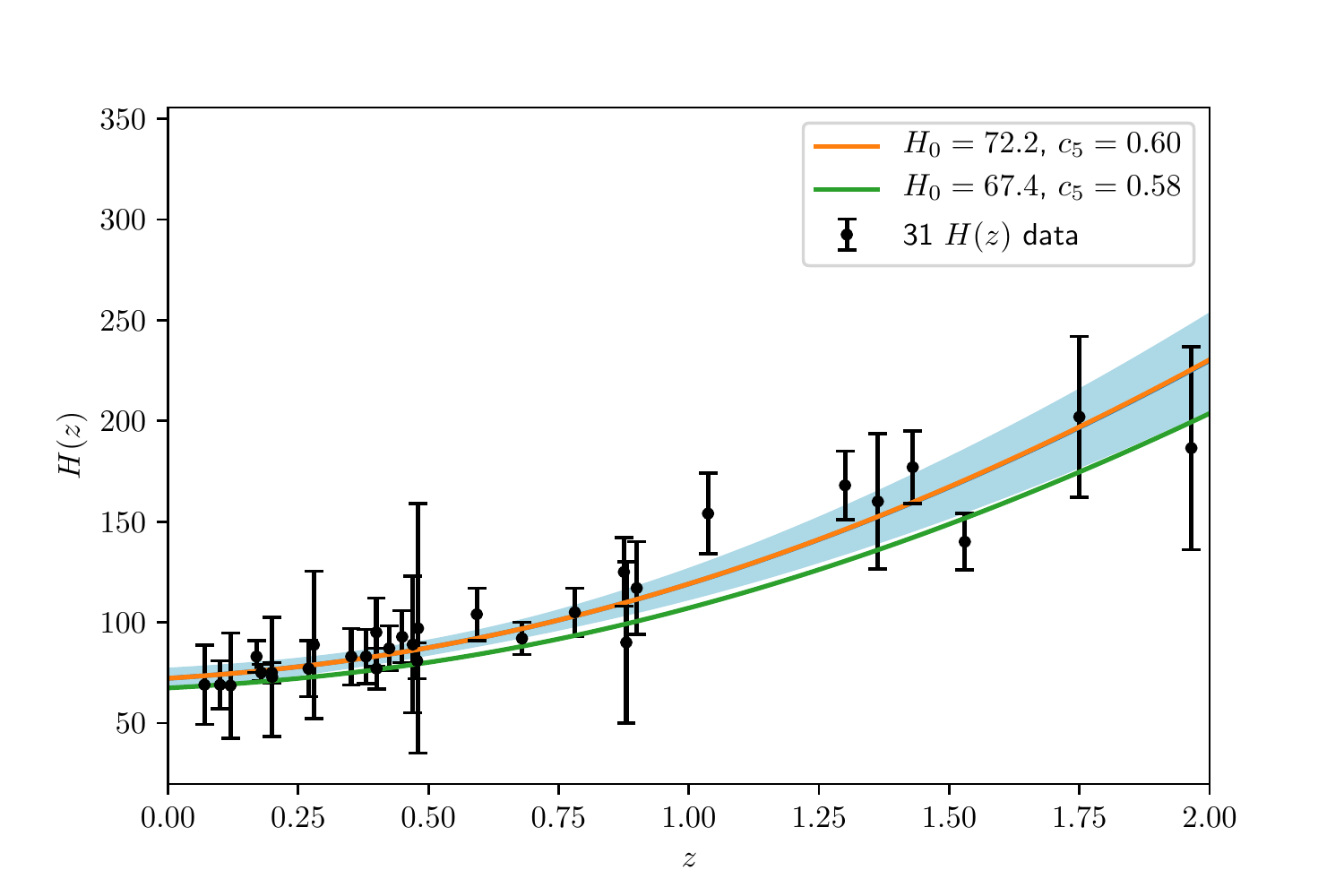}
\caption{Hubble parameter as a function of redshift for the best fit parameters from the 31 $H(z)$ data ($H_0=72.2$ km/s/Mpc, $c_5=0.60$). We also show a curve with $H_0=67.4$ km/s/Mpc, in agreement with Planck data \cite{BAO,Planck} for the Hubble parameter and for a universe age of $13.8$ Gyr, that corresponds to $c_5=0.58$. The blue region corresponds to a 2$\sigma$ (95.4\%) c.l. around the best fit.}
\label{Hz31}
\end{figure*}
 
In order to find the constraints over the free parameters, we have assumed flat priors for $c_5$ and $H_0$ and have sampled the posteriors with the so called Affine Invariant Monte Carlo Markov Chain Ensemble Sampler by \cite{GoodWeare}, which was implemented in {\sffamily Python} language with the {\sffamily emcee} software by \cite{ForemanMackey13}. In order to plot all the constraints on each model, we have used the freely available software {\sffamily getdist}\footnote{{\sffamily getdist} is part of the great Affine Invariant Monte Carlo Markov Chain Ensemble Sampler, {\sffamily COSMOMC} \cite{cosmomc}.}, in its {\sffamily Python} version.

\begin{figure*}[h!]
\centering
\includegraphics[width=.8\linewidth]{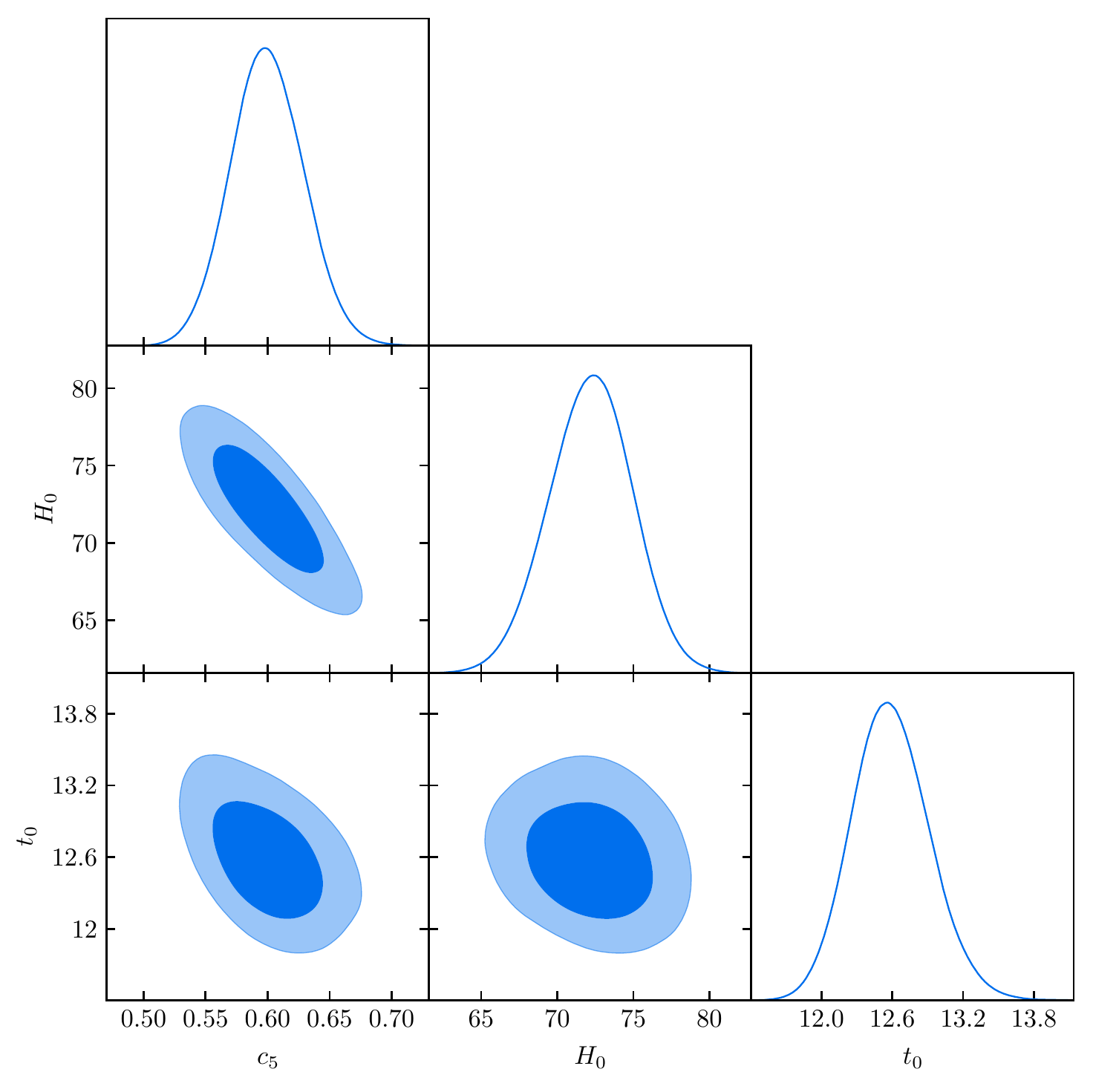}
\caption{Confidence contours from 31 $H(z)$ data analysis of the free parameters of the model, $c_5$ and $H_0$. We also show the constraints over the total age, $t_0$, which is a derived parameter ($H_0$ in km/s/Mpc, $t_0$ in Gyr). The contours correspond to 68\% and 95\% c.l..}
\label{Hz31-triangle}
\end{figure*}

The results of this analysis can be seen in Fig.\ref{Hz31-triangle} and Table \ref{tabHz31}.

\begin{table}[h!]
    \centering
    \begin{tabular} { l  c}
 Parameter &  95\% limits\\
\hline
{\boldmath$c_5            $} & $0.600^{+0.061}_{-0.058}   $\\
{\boldmath$H_0            $} & $72.2^{+5.3}_{-5.5}        $\\
$t_0                       $ & $12.59^{+0.69}_{-0.62}     $\\
\hline
\end{tabular}
\caption{Mean value and 95\% limits of the model parameters. In bold face are the free parameters and $t_0$ is a derived parameter. $H_0$ is in units of km/s/Mpc and $t_0$ in Gyr.}
\label{tabHz31}
\end{table}

%{\color{red} Using the present deceleration factor of  $q_0=-0.377$ (see Eq. \eqref{dfactor}) and the Universe age of $13.8$ Gyr \cite{Planck}, we obtain a Hubble constant of 50.56 km s$^{-1}$ Mpc$^{-1}$ and a constant $c_5 = 0.82$. For those parameters, we can obtain the evolution of the Hubble parameter in terms of the redshift. In Fig. \ref{fig420} we compare our results for the Comoving Hubble parameter with the baryonic acoustic oscillations (BAO) measurements from BOSS DR12 \cite{BAO}, the clustering measurements of quasars from BOSS DR14 \cite{Zarrouk} and the flux-transmission correlations in Ly$\alpha$ forests (BOSS Ly-$\alpha$) \cite{Bautista,Font-Ribera,Planck}, which shows the same general behavior in terms of the redshift.}

\begin{figure*}[h!]
\centering
\includegraphics[width=115mm]{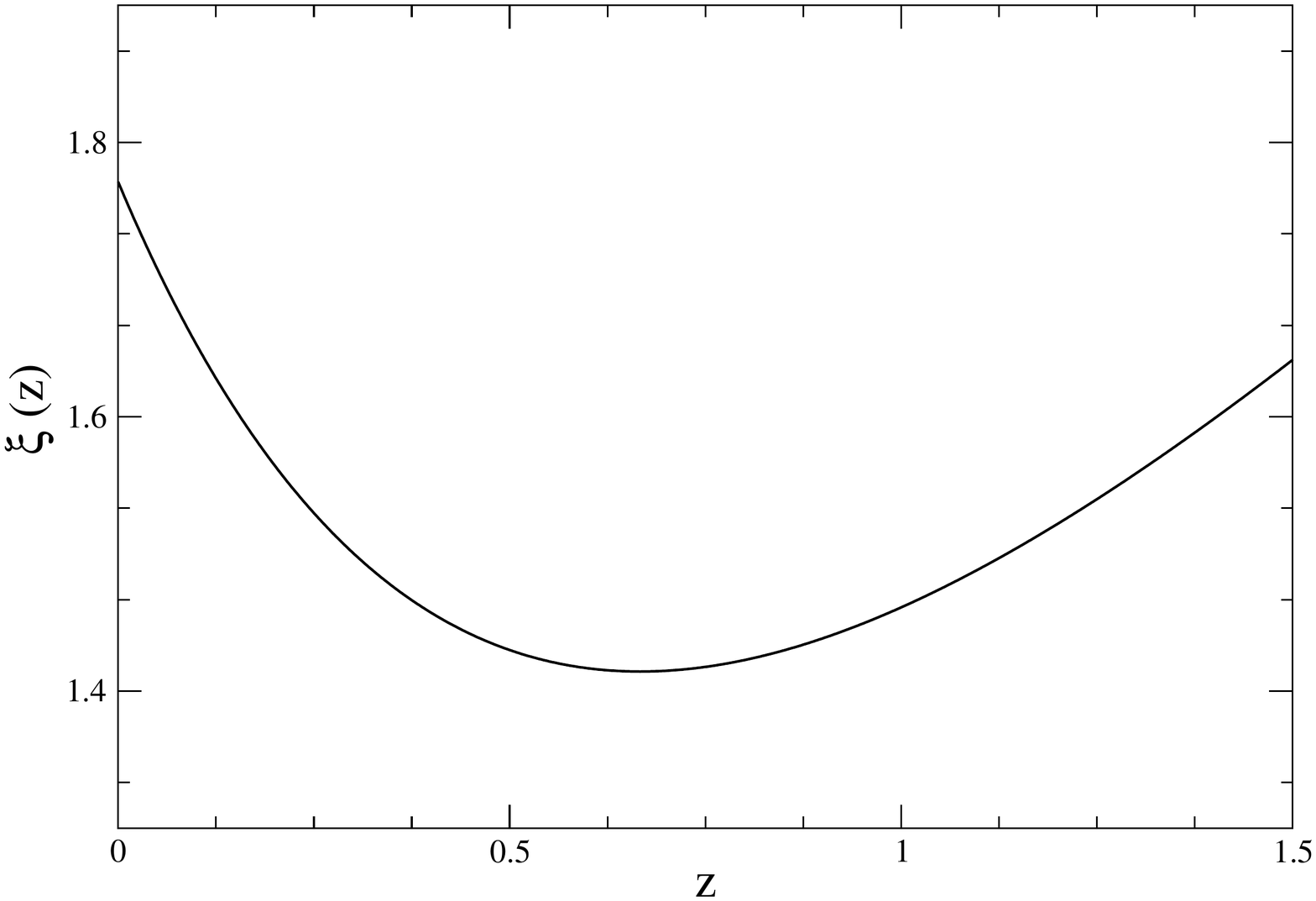}
\caption{Evolution of the extra-dimension scale factor as a function of the redshift in natural units, for $c_5=0.60$ and $c_6=1$.} 
\label{fig41}
\end{figure*}

In Fig. \ref{fig41} we present the scale factor of the extra dimension as function of the redshift, which can be written as

\begin{equation}
\xi(z) = c_5 \, c_6 \, (1+z) \sqrt{1+ \frac{1}{c_5^4 \, (1+z)^4}} \, .
\label{xiz}
\end{equation}

It is interesting to note that the extra-dimension scale factor has a free constant $c_6$, which is not fixed by the cosmological analysis. This happens because the extra dimensional dependence of the cosmological parameters only appears through the fraction $\dot{\xi}/\xi$. This means that, in this model, even if the scale of the extra dimension is very small, the cosmological effects would still be measurable. As a consequence, the extra dimensional lenght scale can be arbitrarly small.

\begin{figure*}[h!]
\centering
\includegraphics[width=115mm]{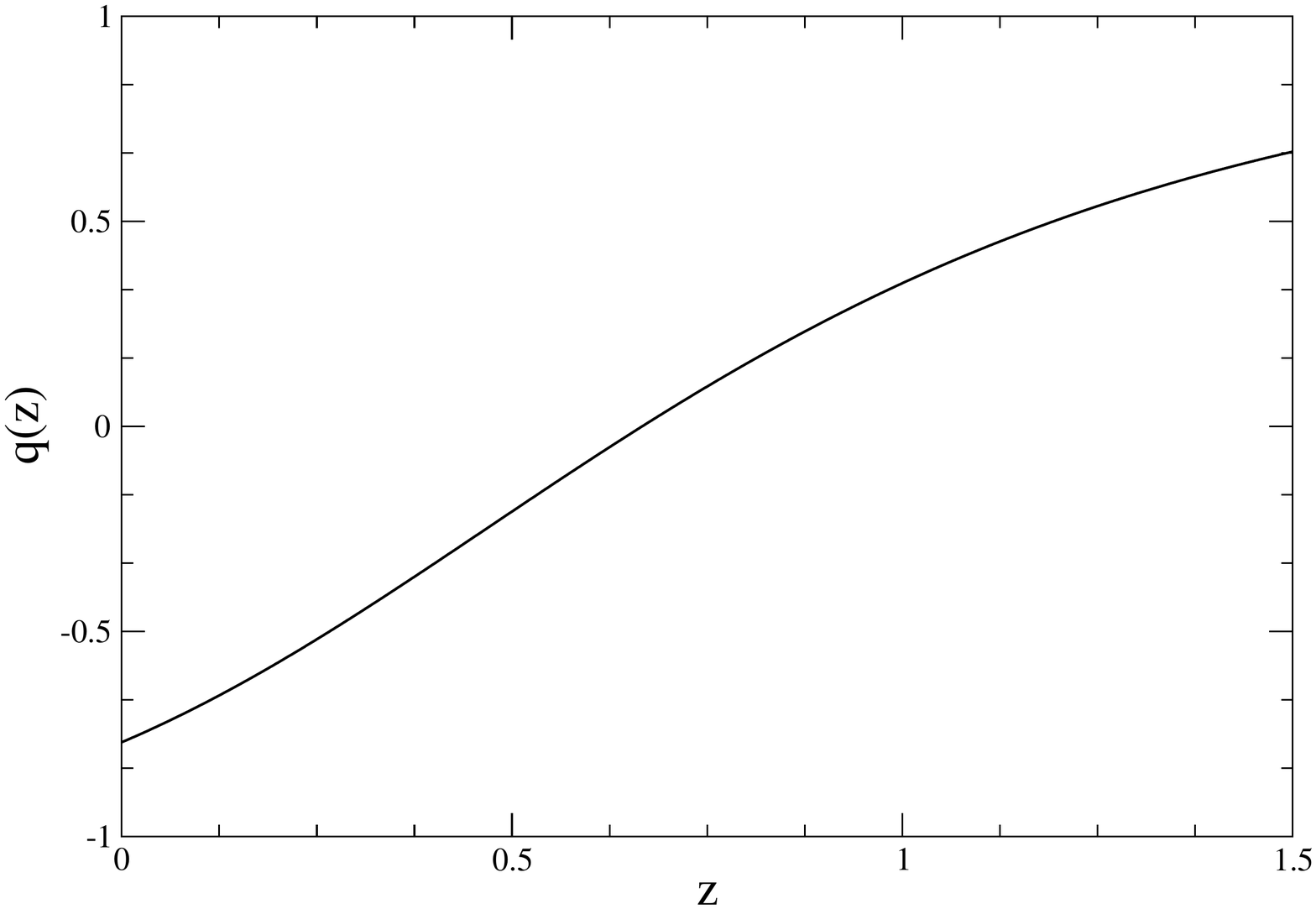}
\caption{Evolution of the universe deceleration parameter as a function of the redshift, for $c_5=0.60$.}
\label{fig44}
\end{figure*}

The deceleration parameter as function of the redshift is given by 
\begin{equation}
q(z) = 1 - \frac{2}{1+c_5^4 \, (1+z)^4} \, ,
\end{equation}
whose behavior can be seen in Fig.\ref{fig44}.  We can see that the model gives an accelerated expansion of the universe ($q<0$) for the present epoch. Also, the solution obtained within the IMM prescription gives a transition from a decelerated to an accelerating universe expansion, as expected from supernova observations, in particular the SN 1997f data photometric observations by the \emph{Hubble Space Telescope} \cite{decelerate2001}. In the analysis of \cite{decelerate2002}, the transition is expected to occur for $z\sim0.5$, which is qualitatively compatible to our model prediction ($z\sim 0.66$).

\begin{figure*}[h!]
\centering
\includegraphics[width=115mm]{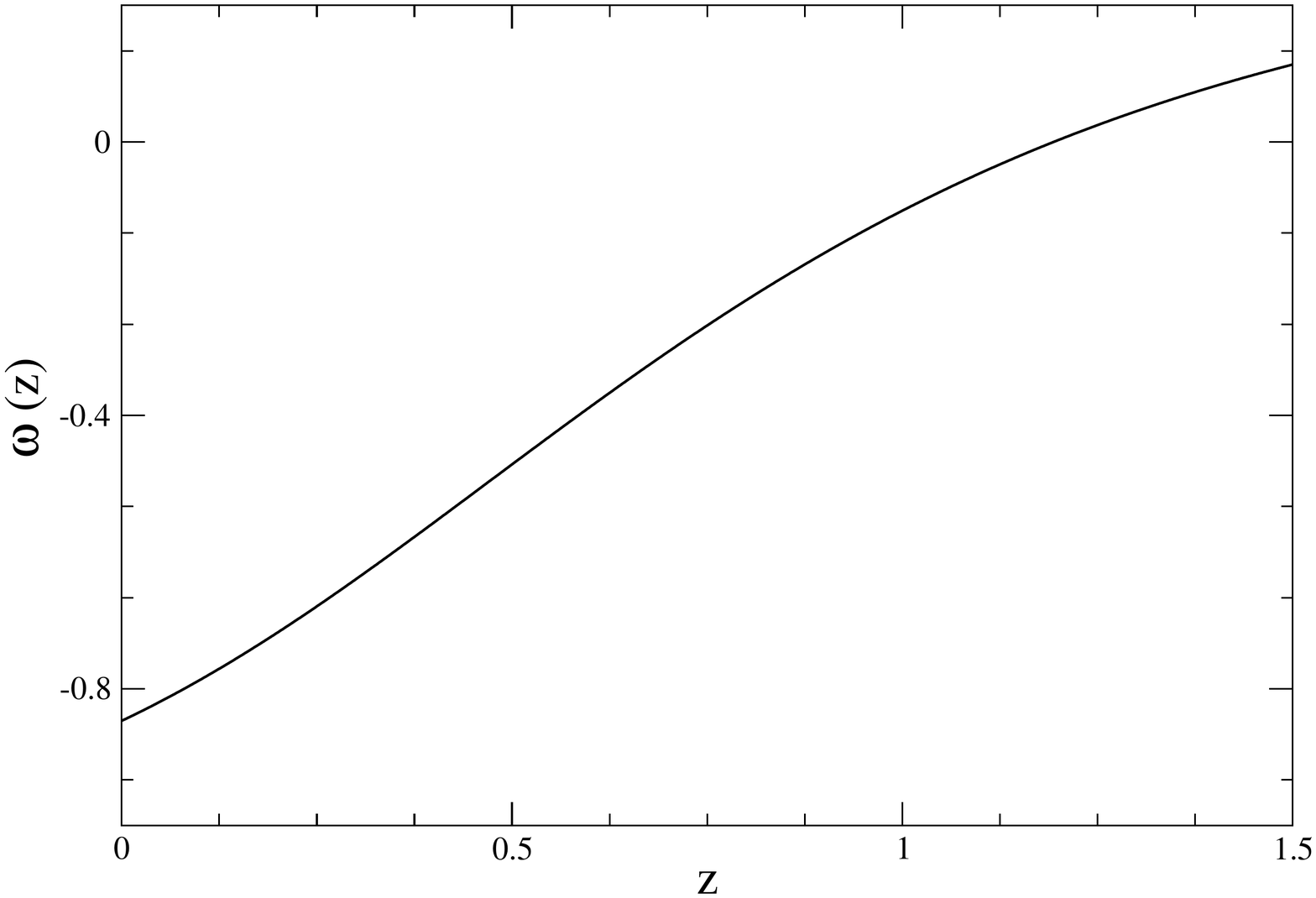}
\caption{Evolution of the EoS parameter as a function of the redshift in natural units, for $c_5=0.60$.}
\label{fig43}
\end{figure*}

The analytical expression for the EoS parameter as function of redshift is 

\begin{equation}
\omega(z) = \frac{1}{3}\left[1-  \frac{4}{1+c_5^4(1+z)^4}\right],
\end{equation}
whose pattern is shown in Fig.\ref{fig43}. It presents the EoS parameter evolution for different epochs of the universe. As expected by the standard cosmological model \cite{PDG}, for recent redshifts the parameter is $<-1/3$ (dark energy era) and for past times the EoS parameter presents a null value, which is compatible with the matter-dominated phase. 

\section{Discussion and Conclusion}\label{sec:dis}

In the present article we have applied the IMM to a general 5D metric with scale factors acting in the usual three space coordinates and in the extra spatial coordinate. We have considered 5D field equations with null and non-null (namely, positive and negative) cosmological constant.

The IMM is a purely geometrical approach in the sense that matter in the 4D universe appears as a manifestation of a 5D empty space. The mechanism which describes that is the collection of the extradimensional dependent terms in the 5D Einstein tensor, which are ``moved'' to the {\it rhs} of the field equations, playing the role of an induced energy-momentum tensor. In this sense, what we have is a realization of the Mach's principle \cite{sciama/1953}-\cite{liu/1995}, which was desired by Einstein for a theory of gravity.

From a quite general approach we have obtained some cosmological features particularly interesting. We have shown in Section \ref{ss:fewcc} that general KK models with null cosmological constant are restrict to a radiation-dominated universe - which evolves as $a\sim t^{1/2}$. Since the matter content obtained from the IMM application has a traceless energy-momentum tensor, this is in agreement with Kaluza's original idea of unifying gravitation and electromagnetism. Also, we have shown that the extra-dimension scale factor yields a negative Hubble parameter for the extra coordinate, i.e., the extra coordinate length compactifies.

In Section \ref{sss:lg0}, we have inserted a positive 5D cosmological constant in the field equations. The approach has led to a cyclic or bouncing universe, i.e., a universe that goes from a collapsing era to an expanding era without displaying the singularity that standard model carries. Bouncing cosmological models are well-known alternatives to inflation and also provide the cosmological perturbations we see today. For a deeper understanding of bouncing cosmological models, besides \cite{steinhardt/2002}-\cite{battefeld/2015}, we refer the reader to \cite{brandenberger/2017}. 

In Section \ref{sss:ll0}, we considered $\Lambda<0$. It is interesting to remark here that usually braneworld models contain negative bulk cosmological constant as a consequence of the appearance of terms $\sim\sqrt{-\Lambda}$ in their Friedmann-like equations \cite{ida/2000,bajc/2000}. 

Our negative cosmological constant model has shown to be able to uniquely describe the radiation, matter and dark energy eras of the universe evolution in a continuous and analytical form, what can be clearly seen, for instance in Fig. \ref{fig6}.

This is a quite non-trivial result. Cosmological models able to describe from a single analytical equation of state the whole history of the universe evolution are rarely obtained in the literature \cite{ms/2016,lima/2013}. References \cite{ms/2016,lima/2013} show cosmological scenarios obtained from $f(R,T^\phi)$ gravity, with $R$ being the Ricci scalar and $T^\phi$ the trace of the energy-momentum tensor of a scalar field $\phi$, and decaying vacuum models, respectively.

This interesting feature is a consequence of the remarkable hyperbolic solution obtained for the scale factor. While we have obtained such a  feature from the model, some other approaches use this solution as a prior {\it ansatz} \cite{chawla/2012,pradhan/2014,mishra/2013,maurya/2017,mishra/2016,nagpal/2019}. This kind of hyperbolic solution also is found in the flat $\Lambda$CDM concordance model, by neglecting radiation. However, in this case, the time dependence of the scale factor is like $a(t)\sim\left[\sinh(t)\right]^{2/3}$, and not $a(t)\sim\left[\sinh(t)\right]^{1/2}$, as we have found here \cite{weinberg/1989,KolbTurner90,LimaBasilakos11,Piattella18}.

Furthermore, from our solution for the matter-energy density (25), it is clear that $\rho\rightarrow constant$ for high values of time, which is also in agreement with standard model. Here, this constant reads $\vert\Lambda\vert/16\pi$, while in standard model it is $\Lambda/8\pi$. The factor $2$ between these energy densities may be due to the fact that the former refers to a 5D space-time, and therefore should be more diluted than a 4D cosmological constant. 
 
Our model also satisfactorily fits the observational data for the experimental measurement of the Hubble parameter, as shown in Section 4.1. The adopted method resulted in $H_0 = 72.2^{+5.3}_{-5.5}$ km/s/Mpc, which is in agreement with the most recent estimate from local observations, $H_0=74.03\pm1.42$ km/s/Mpc \cite{RiessEtAl19}, and, in the limit, also in agreement with the Planck collaboration estimate, $H_0=67.4\pm0.5$ km/s/Mpc \cite{Planck}, in the context of flat $\Lambda$CDM cosmology. As a derived parameter, we have obtained the total age of the Universe as $t_0=12.59^{+0.69}_{-0.62}$ Gyr, which is in agreement with most of age estimates of objects today. Jimenez {\it et al.} \cite{JimenezEtAl19} have obtained, with 22 globular clusters (GC) age estimates from \cite{OMalleyEtAl17}, an weighted average of $t_{GC}=13.0\pm0.4$ Gyr, which is in agreement with our superior limit ($t_0=13.28$ Gyr at 95\% c.l.). Our result is also in agreement with estimates of absolute ages of very-low-metallicity stars, estimated in the range of 13.0 -- 13.535 Gyr, as explained on \cite{JimenezEtAl19} and references therein.

%Further corroborating this result, it is worth mentioning the stress in the measure of $H_0$ obtained experimentally, as treated in \cite {sokol/2019}. The actual value of $H_0$ is still one of the most challenging open problems of physics, with differences between measurements in CMB against local measures, with $67.4\, km/s.Mpc^{-1}$ (CMB) to $H_0 = 76.5\, km/s.Mpc^{-1}$ (Gakaxy brightness fluctuations).
 
Finally, since our unified EoS for the universe was obtained from the AdS$_5$ space-time and $|\Lambda|$ is of the same order of $\Lambda_{4}$, the very small value observed for the cosmological constant has its origin in the energy to create the vacuum in the 5D space-time and is not necessarily related to the vacuum energy of quantum fields in 4D \cite{weinberg/1989}.

\bigskip

\begin{acknowledgements}
M.M.Lapola thanks CAPES, for financial support. PHRSM would like to thank S\~ao Paulo Research Foundation (FAPESP), grants 2015/08476-0 and 2018/20689-7, for financial support. WP thanks CAPES, grant 88881.309870/2018-01 and CNPQ, grants 313236/2018-6 and 438562/2018-6.  M.Malheiro thanks CAPES, CNPq and the  FAPESP thematic project 2013/26258-5 for financial support. R. Valentim would like to thank by Funda\c{c}\~ao de Amparo \`a Pesquisa do Estado de S\~ao Paulo - FAPESP who was supported by thematic project process no. 2013/26258-2 and regular project process no. 2016/09831-0. JFJ is supported by Funda\c{c}\~ao de Amparo \`a Pesquisa do Estado de S\~ao Paulo - FAPESP (Process number 2017/05859-0).
\end{acknowledgements}

\end{document}